\documentclass[amsmath,amssymb,amsfonts]{revtex4}
\usepackage{amsthm,bbm,graphicx,bm}

\newcommand{\ket} [1] {\lvert #1 \rangle}
\newcommand{\bra} [1] {\langle #1 \rvert}

\newcommand{\id}{\mathrm{Id}}

\newtheorem{proposition}{Proposition}

\begin{document}

\title{Exploring scalar quantum walks on Cayley graphs}

\author{Olga Lopez Acevedo}
\thanks{Present address: Department of Physics, Nanoscience Center,
FIN-40014 University of Jyv\"{a}skyl\"{a}, Finland}
\email{lopez@phys.jyu.fi}

\author{J\'er\'emie Roland}
\email{jroland@ulb.ac.be}

\author{Nicolas J. Cerf}
\email{ncerf@ulb.ac.be}

\affiliation{Quantum Information and Communication, Ecole Polytechnique, CP 165/59,
Universit\'e Libre de Bruxelles, B-1050 Brussels, Belgium}

\date{\today}

\begin{abstract}
A quantum walk, \emph{i.e.}, the quantum evolution of a particle 
on a graph, is termed \emph{scalar} if the internal space of 
the moving particle (often called the coin) has a dimension one. 
Here, we study the existence of scalar quantum walks
on Cayley graphs, which are built from the generators of a group.
After deriving a necessary condition on these generators for the existence 
of a scalar quantum walk, we present a general method to express
the evolution operator of the walk, assuming homogeneity of the evolution.
We use this necessary condition and the subsequent constructive method
to investigate the existence of scalar quantum walks on Cayley graphs
of various groups presented with two or three generators.
In this restricted framework, we classify all groups --
in terms of relations between their generators -- that admit 
scalar quantum walks, and we also derive the form of the most general 
evolution operator. 
Finally, we point out some interesting special cases, and extend our study 
to a few examples of Cayley graphs built with more than three generators.
\end{abstract}

\maketitle

\section{Introduction}
Within the traditional (discrete-time) model of quantum computation,
there exist several definitions for a quantum walk on a given graph $G$ over a set of vertices $X$.
In general terms, a quantum walk simulates the evolution of a particle on the
graph with some unitary operator as an evolution operator. The most studied
model, the ``coined model'', defines the evolution operator as the product of two operators.
One is a block diagonal operator which acts non-trivially only on some internal space of the particle, termed the ``coin''. The other operator is a permutation which moves the particle along the edges of the graph. In that model, the most common definition, which has been used for
constructing existing quantum walk algorithms, is the one given by Aharonov \textit{et al.} in \cite{QW1}.
Within this definition, the condition for the existence of such operators is that the graph $G$
is an oriented $d$-regular graph and that it is possible to ``label each directed edge with a
number between one and $d$ such that for each possible label $a$, the directed edges labeled $a$ form a permutation''.

Kendon gave another definition \cite{Kendon03}, which
is valid for non oriented graphs where the coin is of dimension $|X|$, the
number of vertices. In fact both definitions can be seen as particular cases
of the definition given by Montanaro in \cite{Montanaro05}.  Following 
the latter work, a necessary and sufficient condition to define a
coined quantum walk is {\it reversibility}. When a graph is reversible, 
it has a certain number of cycles which may
be used to define a coined quantum walk. The idea is that each cycle is
associated to a basis state of the internal space, using the fact that cycles that do not have a
vertex in common can be associated to the same basis state. The
definition of Aharonov \textit{et al.} requires the graph to be $d$-regular
 (and therefore reversible), with exactly $d$ groups of disjoint cycles including all edges of the graph.
Kendon's definition also requires a reversible graph, and in this case the cycles used
are determined by each directed edge and the corresponding edge with opposite direction.
Therefore, reversibility and the number of cycles in the graph may be used to classify the different
possible definitions in a two operator (coined-shift) model of quantum walks.

If we do not restrict to the coined model (that is, we do not assume that the evolution operator
factorizes into two operators), no sufficient condition is known for
the existence of a unitary evolution operator on a given graph.
Presently, only some necessary conditions are known, one of them being reversibility.
In \cite{QW3}, quantum walks with one
operator and with internal space (coin) are explored. It is shown how the
existence of such walks is related to the properties of the graph and some
examples are given. If the model has an internal space of dimension one, the one-operator
model corresponds to the definition used in \cite{Severini02} and quantum walks
become in this case equivalent to quantum cellular automata as defined by Meyer in \cite{NG}.
Here, we will call this model (one operator and a one-dimensional internal space) a
\emph{scalar quantum walk}.

In \cite{Severini02}, Severini gave a necessary condition
for the existence of a scalar quantum walk on a graph, that he called 
{\it strong quadrangularity}.
The strong quadrangularity implies another condition called 
{\it quadrangularity}. Together with specularity, strong quadrangularity
becomes sufficient for the existence of a unitary evolution on the graph. In
general terms, a graph is called {\it specular} if any two vertices are connected by outgoing edges
to the same set of vertex or do not have any common first neighbor, the same
being required for the ingoing edges. Specularity is equivalent to the condition that the operator may be expressed as the product of
a permutation matrix and a block diagonal matrix, which means that the evolution operator takes
in this case the form of a coined evolution operator 
on a smaller graph~\cite{Severini02-coined}.

In this paper, we will address the general question of determining 
on which Cayley graphs a homogeneous scalar quantum walk can be defined.
First, there is an obvious interest in this question from the point of view 
of quantum algorithmics. Indeed, the answer will help defining gates over graphs, which could then be combined to construct quantum algorithms. 
Moreover, algorithms based on scalar quantum walks in the circuit model
can also be compared in a more natural way to quantum walks 
in continuous-time models of quantum computation, 
such as adiabatic quantum computation.
Apart from these algorithmic applications, there are also true conceptual
differences between scalar and coined quantum walks, which make them both interesting topics of studies on their own. 
In the classical case, note that random walks using
an internal state space exhibit a memory effect,
which makes their classical evolution non-Markovian \cite{CW} in contrast
to random walks with no internal space.
In the quantum case, the behavior of scalar quantum walks
is a more direct consequence of a purely quantum effect, 
namely phase interference, since adding an
internal space necessarily prevents interferences between paths.
Finally, we observe that the efficiency of the algorithms using 
quantum walks varies with
the dimension of the internal space without any obvious reason \cite{QW2},
so that it seems to be important  to investigate both scalar and coined 
quantum walks to solve an algorithmic problem on a given search space.

To identify the Cayley graphs that accept a scalar quantum walk, we will first
verify whether some necessary conditions are satisfied for various graphs.
First, we know that all Cayley graphs are reversible. The other known 
conditions are quadrangularity and strong quadrangularity.
While the condition of strong quadrangularity depends on the number of subsets
of the vertex set of the graph, and thus requires an exponential time to be
checked, the quadrangular condition depends on the number of pairs of vertices
and may be checked in polynomial time. We will thus use quadrangularity
as a necessary condition. This paper is organized as follows. In Section~\ref{conditions}, we will derive the quadrangular condition
and adapt it to homogeneous quantum walks on Cayley graphs. 
In Section~\ref{operator}, we will present a general method 
allowing us to determine whether a homogeneous scalar quantum walk exists
on a given Cayley graph and, if so, to explicitly derive its evolution operator.
In Section~\ref{applic}, we will apply the above necessary condition
and constructive method in order to find new scalar quantum walks.  
We will consider a variety of Cayley graphs, built from two, three,
and occasionally more than three generators.

\section{Conditions on the existence of scalar quantum walks on Cayley graphs\label{conditions}}
\subsection{Quadrangularity for general graphs}
Let $G(X,E)$ be a graph with vertex set $X$ and edge set $E$. A scalar quantum
walk is a model of the evolution of a particle on the graph $G$. The state of the
particle is described by a unit vector in the Hilbert space $H=\ell^2(X)$ and we will
use the set of states $\{ |x \rangle \}_{x \in X}$ as a basis for $H$.
The evolution operator $W$ is a
unitary operator on $H$. The evolution of $|\psi_t\rangle$, the state
describing the position of the particle at time $t$, is given by the following equation:
\begin{equation}
\ket{\psi_t} = W^t \ket{\psi_0},
\end{equation}
where $|\psi_0\rangle$ is a given initial state.
We denote as $W_{x,y}$ the matrix elements $\bra x W \ket y$.
A necessary condition for the existence of a scalar quantum walk on a graph
is called quadrangularity. 
It was first presented in \cite{Severini02}, and can also be
obtained as a consequence of the analysis presented in \cite{QW3}.
Let us recall it here.

\begin{proposition}\label{quadrangularity-general}
Given a graph $G(X,E)$, if a scalar quantum walk on this graph exists then for
each pair of different edges of the form $(x,z)$ and $(y,z)$ there exists
another pair of different edges of the form $(x,z')$ and $(y,z')$.
\end{proposition}

\begin{proof}
The equation $W^\dagger W=\id_H$, where $W^\dagger$ is the conjugate transpose
of $W$ and $\id_H$ is the identity on $H$, is equivalent to the set of equations:
\begin{equation}\label{unitary1}
\begin{cases} \sum_{z \in X} \overline W_{z,x} W_{z,y}=0 & \forall\ x,y \in X, x \not = y\\
\sum_{z \in X} \overline W_{z,x} W_{z,x}=1 & \forall\ x \in X\\
\end{cases}
\end{equation}
And the matrix equation $W W^\dagger=\id_H$ is equivalent to
\begin{equation}\label{unitary2}
\begin{cases} \sum_{z \in X} W_{x,z} \overline W_{y,z}=0 & \forall\ x,y \in X, x \not = y\\
\sum_{z \in X} W_{x,z} \overline W_{x,z}=1 & \forall\ x \in X\\
\end{cases}
\end{equation}
where $\overline \alpha$ denotes the complex conjugate of $\alpha$.
Suppose that the sum in the first type of equation contains only one term for some couple $(x,y)$.
Then, the only possibility for the condition to be satisfied is to have one of the two coefficients equal to
zero. This is equivalent to modifying the graph, so that in this case, there is no scalar quantum
walk for the original graph $G$.
\end{proof}

This proposition can also be reformulated as follows:\\
{\it Given a graph $G(X,E)$, if a scalar quantum
walk on this graph exists then all consecutive edges in the graph with different
orientations belong to at least one closed path of length four with
alternating orientations}.

\subsection{Quadrangularity for Cayley graphs}
A Cayley graph is a graph constructed using the relations of a group.
Given a group $\Gamma$ and a set $\Delta$ of elements generating the group (all elements
in $\Gamma$ may be obtained by multiplication of elements and inverse elements of
the generating set), the Cayley graph $C_{\Delta}(\Gamma)=(X,E)$ is defined by 
\begin{align}
X&=\Gamma\\
E&=\{(x,x \delta):x \in \Gamma,\delta \in \Delta \}
\end{align}

By this definition it will be natural to suppose that the coefficients of the
evolution operator depends only on the generator used and not on the starting
vertex, which can be interpreted as a homogeneity condition
\begin{equation}\label{homogeneity}
W_{x \delta,x}=W_{\delta}.
\end{equation}
In this case, the evolution operator may be written
\begin{equation}
W=\sum_{\delta\in\Delta} W_\delta U_\delta,
\end{equation}
where $U_\delta=\sum_x \ket{x\delta}\bra{x}$ is the unitary operator that
right-multiplies each group element by $\delta$.
The unitarity equations (\ref{unitary1}-\ref{unitary2}) becomes
\begin{equation}\label{unitaryCayley}
 \sum_{\delta_1 \delta_2^{-1}=u} {\overline{W}}_{\delta_1}
  W_{\delta_2} = \delta_{u=e}\qquad \forall\ \delta_1,\delta_2 \in \Delta,
\end{equation}
where $\delta_{u=e}=1$ if $u$ equals the identity element $e$ in the group $\Gamma$, and $0$ otherwise.

For quantum walks over a Cayley graph,
Proposition~\ref{quadrangularity-general} implies a necessary condition over the elements of
the generating set $\Delta$:
\begin{proposition}\label{quadrangularity-cayley}
Given a group $\Gamma$ and a set $\Delta$ of elements generating $\Gamma$,
a necessary condition for the existence of a scalar quantum walk of the form
$W=\sum_{\delta\in\Delta} W_\delta U_\delta$ on the Cayley graph
$C_{\Delta}(\Gamma)$
is that for all couples $(\delta_1,\delta_2)$ of
different elements of $\Delta$, there exists at least one different couple
$(\delta_3,\delta_4)$ such that  $\delta_1 \delta_2^{-1}=\delta_3
\delta_4^{-1}$.
\end{proposition}
\begin{proof}
From Equation (\ref{unitaryCayley}), we see that when the necessary condition is
not satisfied, at least one of the coefficients $W_i$ has to be zero. This is
equivalent to modifying the graph (to removing the corresponding edges labeled
with~$i$).
\end{proof}

In principle, it is impossible to solve the general problem of determining 
and classifying all groups obtained by the procedure of taking the quotient 
of a free group. This result is a consequence of the existence 
of finitely presented groups for which the word problem is unsolvable 
(see Theorem $7.2$ in \cite{LS77}).
For this reason, it is probably hopeless to try solving 
the problem of determining and classifying all groups 
that verify the quadrangularity condition.

Here, we will instead consider groups with a small number of generators 
and build all possible minimal sets of relations verifying the quadrangularity condition. For the resulting
groups, we will completely solve the system of equations~(\ref{unitaryCayley})
arising from the unitarity condition, and thus derive 
the most general evolution operator for a scalar quantum walk 
on the associated Cayley graphs. The general method to solve this problem is exposed in the next Section.

\section{Solving the unitarity condition for the evolution operator\label{operator}}

Let us suppose we want to build a scalar quantum walk on a Cayley graph verifying
the necessary condition from Proposition~\ref{quadrangularity-cayley}.
When the homogeneity condition (\ref{homogeneity}) is imposed,
the number of variables describing the evolution operator decreases considerably since it is reduced
to $d=|\Delta|$, the number of elements in the generating set $\Delta$.
The number of equations in the system (\ref{unitaryCayley}) is also reduced,
it is upper bounded by the number of different pairs of generators, that is, $\frac{d(d-1)}{2}$.
Each equation in this system may be transformed into a bilinear equation
\begin{eqnarray}
\bm{v}^\dagger P \bm{v} = 0,\label{bilinear}\\
\bm{v}^\dagger \bm{v} = 1,\label{normalization}
\end{eqnarray}
where $P$ is a $d\times d$ matrix with entries in $\{0,1\}$
depending on the graph $C_\Delta(\Gamma)$ and
$\bm{v}\in\mathbbm{C}^d$ is a column vector having as components the elements $W_i$ of
the evolution operator.
More precisely, if we define as
\begin{equation}
\bm{e}_1=\left( \begin{matrix} 1\\ 0 \\ \vdots \\ 0\end{matrix} \right),
\bm{e}_2=\left( \begin{matrix} 0\\ 1 \\ \vdots \\ 0\end{matrix} \right),
\cdots,
\bm{e}_d=\left( \begin{matrix} 0\\ 0 \\ \vdots \\ 1\end{matrix} \right)
\end{equation}
the vectors in the standard basis of $\mathbbm{C}^d$, we have
$\bm{v}=\sum_{i=1}^d W_i \bm{e}_i$.
Let $\{\lambda_1,\dots, \lambda_d\}$ be the $d$ eigenvalues of $P$
(with algebraic multiplicities). If the geometric multiplicities of these eigenvalues
equal their algebraic multiplicities, we may build an orthonormal basis
$\{\bm{v}_1, \dots, \bm{v}_d\}$ out of eigenvectors of $P$.
Developing $\bm{v}$ in this basis, we may write
$\bm{v} = \sum_{j=1}^d \alpha_j \bm{v}_j$
so that Equation (\ref{bilinear}) becomes
\begin{equation} \label{linear1}
\sum_{j=1}^d \lambda_j | \alpha_j |^2=0,
\end{equation}
whereas the normalization equation (\ref{normalization}) is simply
\begin{equation} \label{linear2}
\sum_{j=1}^n | \alpha_j |^2 =1.
\end{equation}
The relation between the new coefficients $\alpha_j$ and the evolution operator
coefficients $W_i$ is then given by
\begin{equation}
W_i = \sum_j  \alpha_j\ \bm{e_i}^\dagger\bm{v_j}.
\end{equation}

In the next section, we will use this approach to build scalar quantum walks on groups with
few generators, or prove that no such walk exists.

\section{Application}
\label{applic}

Let us consider groups denoted as $\Gamma=\langle \Delta|R\rangle$,
using the standard free group representation where
$\Delta$ is a set of generators for the group $\Gamma$
and $R$ is a set of relations among these generators,
and determine whether the associated Cayley graph admits a scalar
quantum walk.

\subsection{Groups finitely presented with two generators\label{two-generators}}
The simplest case is a group finitely generated with two generators.
We denote the set of generators as $\Delta=\{x,y\}$. There are only two possible couples $(x,y)$
and $(y,x)$ and therefore the only relation which makes the necessary condition satisfied is $x y^{-1}=y x^{-1}$. Let us consider the corresponding group
$\Gamma=\langle x,y|x y^{-1}=y x^{-1}\rangle$, and check whether there is a scalar
quantum walk on the associated Cayley graph  $C_\Delta(\Gamma)$.
The unitarity equations (\ref{unitaryCayley}) become
\begin{equation}
\begin{cases} \overline W_x W_y+ \overline W_y W_x=0\\
\overline W_x W_x+ \overline W_y W_y=1,
\end{cases}
\end{equation}
which can be written as bilinear equations (\ref{bilinear}-\ref{normalization})
with $\bm{v}=\left(
  \begin{matrix} W_x\\W_y \end{matrix} \right)$
and $P=\left(
  \begin{matrix} 0 & 1\\1 & 0 \end{matrix} \right)$.
Developing $\bm{v}$ in the basis formed by the eigenvectors of $P$, that is
$\bm{v}_1=\frac{1}{\sqrt{2}}\left(
  \begin{matrix} 1\\1 \end{matrix} \right)$ and
$\bm{v}_2=\frac{1}{\sqrt{2}}\left(
  \begin{matrix} 1\\-1 \end{matrix} \right)$,
with respective eigenvalues $\lambda_1=1$ and $\lambda_2=-1$, we see that
Equations~(\ref{linear1}-\ref{linear2}) become
\begin{equation}
\begin{cases}
|\alpha_1|^2 -|\alpha_2|^2=0\\
|\alpha_1|^2 + |\alpha_2|^2=1,
\end{cases}
\end{equation}
so that the solution is up to a global phase
\begin{equation}\label{sol2gen}
\begin{cases}
W_x= \cos \phi\\
W_y= i \sin \phi
\end{cases}
\end{equation}
for $\phi \in [ 0,2\pi]$.
Thus, there exists a scalar quantum 
walk on the Cayley graph $C_\Delta(\Gamma)$ for the group considered here.

Interestingly, this is also true for all groups finitely presented with
two generators satisfying $x y^{-1}=y x^{-1}$, since adding other types of relation will not modify the system of equations~(\ref{unitaryCayley}).
In particular, we can add some relations to the previous infinite group in order to make
it finite. Hence, we show that there is a scalar quantum walk for the
dihedral group $D_{n}=\langle x,y|x^{n}=e,y^2=e,(xy)^2=e\rangle$,
with generating set $\Delta=\{x,y\}$. The evolution operator coefficients are also given by Eqs.~(\ref{sol2gen}), and the
corresponding Cayley graph is both strongly quadrangular and specular
(it was first presented in \cite{Severini02}).

Finally, let us note that the hypercube of dimension 2 can be obtained as
the Cayley graph of the $n=2$ case of $D_{n}$, that is,
$D_2=\langle x,y|x^2=e,y^2=e,xy=yx\rangle$, with $\Delta=\{x,y\}$.
We conclude that the hypercube of dimension 2
therefore also accepts a scalar quantum walk.

\subsection{Groups finitely presented with three generators\label{three-generators}}
\begin{figure}[htb]
\begin{center}
\includegraphics[height=3cm]{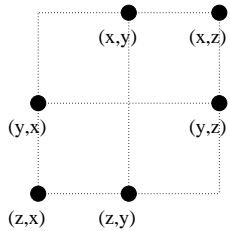}  
\end{center}
\caption{Representation of the relations between generators as a grid.
A relation $\delta_i\delta_j^{-1}=\delta_k\delta_l^{-1}$ will be represented as
 an edge between vertices $(\delta_i,\delta_j)$ and $(\delta_k,\delta_l)$.\label{reduction}}
\end{figure}

In the case of three generators there are six possible couples, and then a
large number of groups, in principle $6^5$.
We will use the following representation 
in order to simplify the counting of the
possibles graphs verifying the necessary condition. We associate each pair of
generators $(\delta_i,\delta_j)$ to the vertex element $M_{i,j}$ of a $3
\times 3$ grid $M$ (see Fig.~\ref{reduction}). A relation $\delta_i\delta_j^{-1}=\delta_k\delta_l^{-1}$
may then be represented as an edge between vertices $(\delta_i,\delta_j)$
and $(\delta_k,\delta_l)$.

From the necessary condition, we know that the grids corresponding to a valid graph are those where all vertices are connected with
at least another vertex. Note that there is no element in the diagonal, since
there are no couples of the form $(\delta_i,\delta_i)$. Moreover, the different edges are not independent, since $\delta_i\delta_j^{-1}=\delta_k\delta_l^{-1}$ is equivalent to
$\delta_j\delta_i^{-1}=\delta_l\delta_k^{-1}$
(this implies a symmetry with respect to the diagonal axis), while by transitivity the relations
$\delta_i\delta_j^{-1}=\delta_k\delta_l^{-1}$ and $\delta_k\delta_l^{-1}=\delta_m\delta_n^{-1}$
imply $\delta_i\delta_j^{-1}=\delta_m\delta_n^{-1}$
We also see that not all edges are
interesting since a vertical or an horizontal edge implies a relation of
the form $\delta_i=\delta_j$ and thus a trivial reduction of the number of generators.

\begin{figure}[htb]
\begin{center}
 \includegraphics[height=2.5cm]{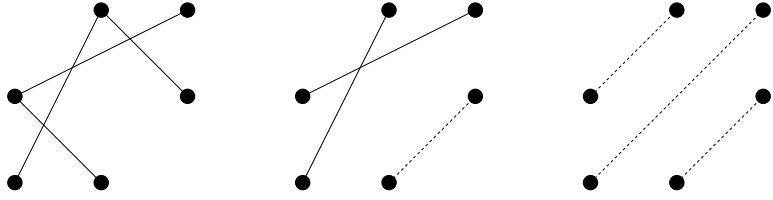}  
\end{center}
\caption{Representation of the three sets of relations implying no reduction.
Valid sets of relations are those for which any vertex is connected to at least one
other vertex. Other sets of relations either reduce to one of the three sets represented here
(possibly up to a permutation of the generators) or imply a reduction of the number of generators.
\label{groups-without-reduction}}
\end{figure}

By a straightforward (but tedious) inspection of all grids,
it can be shown that among the $6^5$ groups finitely presented 
with three generators, there are only three non-equivalent groups 
(without trivial reduction) that satisfy the
necessary condition (see Fig.~\ref{groups-without-reduction}).
There remains to check whether there actually exists a scalar quantum walk
on the Cayley graph built from these three groups:
\begin{itemize}
\item [{\it i})]
$\Gamma=\langle x,y,z|xy^{-1}=yz^{-1},xy^{-1}=zx^{-1} \rangle$ and $\Delta=\{x,y,z\}$\\
Note that the group $\Gamma$ may equivalently be written as
  $\langle x,y,z|z=xy^{-1}x,(xy^{-1})^3=e \rangle$.
The unitarity condition equations (\ref{unitary1}-\ref{unitary2}) become
\begin{equation}
\begin{cases} \overline W_x W_y+ \overline W_z W_x +\overline W_y W_z =0\\
\overline W_x W_x+ \overline W_y W_y + \overline W_z W_z =1,
\end{cases}
\end{equation}
which can be written as a bilinear hermitian form with $\bm{v}=\left(
  \begin{matrix} W_x\\W_y\\W_z \end{matrix} \right)$
and $P=\left(
  \begin{matrix} 0 & 1 & 0\\ 0 & 0 & 1\\1 & 0 & 0 \end{matrix} \right)$.
Using the expansion of $\bm{v}$ in terms of the eigenvectors of $P$, the system reduces to
\begin{equation}
\begin{cases}
|\alpha_1|^2 + e^{i \frac{2\pi}{3}}|\alpha_2|^2 + e^{-i \frac{2\pi}{3}}|\alpha_3|^2=0\\
|\alpha_1|^2 + |\alpha_2|^2 + |\alpha_3|^2=1,
\end{cases}
\end{equation}
so that the solution is (up to a global phase)
\begin{equation}\label{sol3gen}
\begin{cases}
W_x= \frac{1}{3}(1+ e^{i \phi_1} + e^{i \phi_2})\\
W_y= \frac{1}{3}(1+ e^{i \frac{2\pi}{3}}e^{i \phi_1} + e^{-i \frac{2\pi}{3}}e^{i \phi_2})\\
W_z= \frac{1}{3}(1+ e^{-i \frac{2\pi}{3}}e^{i \phi_1} + e^{i \frac{2\pi}{3}}e^{i \phi_2}).
\end{cases}
\end{equation}
\item [{\it ii})]
$\Gamma=\langle x,y,z| xy^{-1}=zx^{-1},yz^{-1}=zy^{-1} \rangle$ and $\Delta=\{x,y,z\}$\\
Using a similar method, we find the following general solution:
\begin{equation}\label{sol3gen-2}
\begin{cases}
W_x= \cos\phi \\
W_y= \frac{1+i(-1)^q}{2} \sin\phi \\
W_z= -\frac{1-i(-1)^q}{2} \sin\phi,
\end{cases}
\end{equation}
for $\phi \in [ 0,2 \pi ]$ and $q$ any integer.
\item [{\it iii})]
$\Gamma=\langle x,y,z| xy^{-1}=yx^{-1},xz^{-1}=zx^{-1}, yz^{-1}=zy^{-1}
  \rangle$ and $\Delta=\{x,y,z\}$\\
We obtain that there is no solution
(the proof is also based on the same method as above).
\end{itemize}

\begin{figure}[htb]
\begin{center}
 \includegraphics[height=2.5cm]{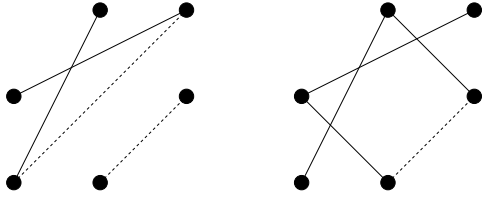}
\end{center}
\caption{Representation of the two sets of relations implying
reductions of the number of generators. In the first case, we obtain $x=y$, while in the second,
we have $x=y=z$. Note that other sets of relations may reduce to these cases, possibly up to
some permutation of the generators.\label{groups-with-reduction}}
\end{figure}

If we add more relations of the type $\delta_i\delta_j^{-1}=\delta_k\delta_l^{-1}$
to these groups, reductions appear and we end up with one of the following groups
with less than three generators (see Fig.~\ref{groups-with-reduction}):
\begin{itemize}
\item [{\it iv})]
$\Gamma=\langle x,y | x y^{-1}=y x^{-1} \rangle$ and $\Delta=\{x,y\}$\\
This group with two generators has been presented in Section~\ref{two-generators}
and admits a scalar quantum walk.
\item [{\it v})]
$\Gamma=\langle x \rangle$ and $\Delta = \{ x \}$\\
The free group with one generator admits a trivial scalar quantum walk, 
which acts as a shift operator
on an infinite line. Adding a relation $x^n=e$, we obtain the cyclic group
$C_n=\langle x| x^n=e\rangle$, and the associated Cayley graph becomes a cycle,
which thus also accepts such a trivial scalar quantum walk.
\end{itemize}

Of course, there exist other relations beyond these of the type
$\delta_i\delta_j^{-1}=\delta_k\delta_l^{-1}$, so that we may build further groups by adding
other relations. Nonetheless, as long as these additional relations do not imply relations
of the previous type nor trivial reductions, they will have no influence on the
possible existence of a scalar quantum walk on the associated Cayley graphs.

For instance, adding relations $x^2=e$ and $y^2=e$ 
to the above group ({\it i}),
we obtain the symmetric group $S_3=\langle x,y|x^2=e,y^2=e,(xy)^3=e\rangle$ with two generators. This is
the group of permutations of three elements, $x$ corresponding 
to a permutation of the first two elements, $y$ to a permutation of the last two elements, and $xyx$ to a permutation of the first and third
elements. Taking $\Delta=\{x,y,xyx\}$, the corresponding Cayley graph therefore admits a scalar quantum walk, with evolution operator coefficients given by Eq.~(\ref{sol3gen}).

Similarly, if we add the relation $y^2=x$ to the group ({\it ii}),
the set of relations imply $y^3=z$ and $y^4=e$, so that we obtain the cyclic graph
$C_4=\langle y|y^4=e\rangle$. Since we use as generators all the group elements
except the identity $e$, that is, $\Delta=\{y,y^2,y^3\}$, the Cayley graph reduces to the complete
graph over four vertices, which thus admits a scalar quantum walk with evolution operator coefficients given by Eq.~(\ref{sol3gen-2}).

Finally, if we impose the additional relations $x^2=y^2=z^2=e$ to the group ({\it iii}) without reduction,
the associated Cayley graph becomes the hypercube in dimension $3$. It follows that even though
there exists a scalar quantum walk on the hypercube in dimension $2$, as we have seen above, there is no such walk in dimension $3$.

\subsection{Other examples}
\begin{itemize}
\item [{\it vi})]
Cayley graphs of the cyclic group $C_n=\langle a| a^n=e\rangle$.\\
We have already found a way to define a scalar quantum walk on a cycle, which is the simplest Cayley graph of the cyclic group $C_n$. Here, we show how to build
homogeneous scalar quantum walk on more general Cayley graphs 
of the cyclic group (see also \cite{Severini03}).
The elements of $C_n$ may be written as $\{a^i|i=0,\dots,n\}$.
The necessary condition implies that if $a^i$ and $a^j$ are taken as generators
to build the Cayley graph, we must also take some elements $a^k$ and $a^l$
such that $(i-j)-(k-l) \mod n=0$.

In particular, if $n$ is a multiple of $d$, this condition is satisfied if the generator set,
with $d$ elements, is taken as $\Delta = \{a^{j \frac{n}{d}+l}|j=0,\dots,d-1 \}$,
where $l$ is some integer.
The unitarity condition equations (\ref{unitary1}-\ref{unitary2})
become
\begin{equation}
\sum_{j=0}^{d-1}\overline W_j W_{(j+i)\!\!\!\mod n} = \delta_{i,0}
\end{equation}
for $i = 0,\dots,d-1$ and the solution is
\begin{equation} 
W_j=\frac{1}{d} \sum_{k=0}^{d-1}e^{i \theta_k} e^{i \frac{2 \pi k
    j}{d}}.
\end{equation} 
Hence, there are $d-1$ real parameters that define the evolution operator
for this family of Cayley graphs of the cyclic groups. This is not exhaustive for cyclic groups and
there are also scalar quantum walks on Cayley graphs
outside this family, we have already seen one example, the complete graph
over $4$ vertices, and another example will be given below, in the context of Johnson graphs.

The $d=n$ special case
corresponds to the complete graph with self-loops (the self-loops come from the fact that
we also take as generator the identity $e$, which maps any group element to itself).
This graph trivially accepts scalar quantum walks since any unitary matrix without zero entries may be taken as a diffusion operator (note that here we give the general form of \emph{homogeneous}
scalar quantum walks). In particular, taking $\theta_0=\pi$ and $\theta_k=0$ for all $k\neq 0$
yields the usual diffusion operator from Grover's algorithm, with $W_0=1-2/n$ and $W_k=-2/n$ for all $k\neq 0$.
The $d=1$ special case reduces to the cycle, that we have already seen
above in the context of the free group with one generator.

\item [{\it vii})]
The Johnson graphs $J(n,k)$.\\
The Johnson graph $J(n,k)$ is defined as the graph having as vertices the $C^k_n$ subsets
of size $k$ of a set of $n$ elements, and such that two subsets are linked by an edge
if they have $k-1$ elements in common. Let us note that $J(n,n-k)=J(n,k)$.

All Johnson graphs verify the quadrangularity condition. Numerically, we have
obtained that the Johnson graphs $J(n,k)$ for small values of $n$ and $k$ also verify the strong
quadrangular condition. Whether they admit a scalar quantum walk in general is an open question.
However, our method may help to solve this problem on a case by case basis.
Indeed, while until now we have built Cayley graphs from groups, we may take the opposite point of view and try to build a group that 
accepts a given graph as Cayley graph.
For instance, the graph $J(4,2)$ can be described as a Cayley graph of the cyclic group
of order six $C_6=\langle\delta_1|\delta_1^6=e\rangle$. Defining $\delta_2=\delta_1^2$,
the Cayley graph, represented in Fig.~\ref{johnson}, is then constructed with the generating set $\Delta=\{\delta_1,\delta_1^{-1},\delta_2,\delta_2^{-1}\}$.
The equations are thus
\begin{equation}
\begin{cases}
\overline W_{\delta_2} W_{\delta_1^{-1}}+\overline W_{\delta_2^{-1}}
W_{\delta_1}+\overline W_{\delta_1} W_{\delta_2^{-1}}+\overline
W_{\delta_1^{-1}} W_{\delta_2}=0\\ 
\overline W_{\delta_2} W_{\delta_1}+\overline W_{\delta_1^{-1}} W_{\delta_2^{-1}}=0\\
\overline W_{\delta_1} W_{\delta_1^{-1}}+\overline W_{\delta_2^{-1}} W_{\delta_2}=0\\
\overline W_{\delta_1} W_{\delta_1}+\overline W_{\delta_1^{-1}}
W_{\delta_1^{-1}}+\overline W_{\delta_2} W_{\delta_2}+\overline
W_{\delta_2^{-1}} W_{\delta_2^{-1}}=1.
\end{cases}
\end{equation}
Using the previous method, we show that the general solution is
\begin{equation}
\begin{cases}
W_{\delta_1}=\frac{1}{2}e^{i \phi}\\
W_{\delta_1^{-1}}=\frac{1}{2}\\
W_{\delta_2}=\frac{(-1)^q}{2}\\
W_{\delta_2^{-1}}=\frac{(-1)^{q+1}}{2}e^{i \phi}
\end{cases}
\end{equation}
for $\phi \in [ 0,2 \pi ]$ and $q$ any integer.

\begin{figure}[htb]
\begin{center}
 \includegraphics[height=3.5cm]{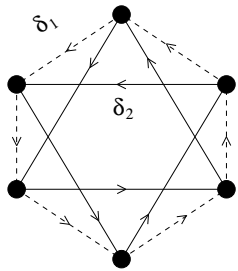}
\end{center}
\caption{Johnson graph $J(4,2)$, which may be seen as the Cayley graph of the cyclic group
$C_6=\langle\delta_1|\delta_1^6=e\rangle$ with generating set $\Delta=\{\delta_1,\delta_1^{-1},\delta_2,\delta_2^{-1}\}$, where $\delta_2=\delta_1^2$.
Note that we have only represented the oriented edges corresponding to $\delta_1$ and
$\delta_2$, and not their inverses, corresponding to $\delta_1^{-1}$ and $\delta_2^{-1}$.
\label{johnson}}
\end{figure}

Going beyond $J(4,2)$, we have also tested
numerically whether Johnson graphs verify the strong quadrangularity
condition. We have obtained that, at least within a
computationally achievable range of values $n$ and $k$, they
always verify the strong quadrangularity condition, 
except for the pathological case $J(3,1)$ (and consequently $J(3,2)$)
which corresponds to the complete graph over three vertices 
(without self-loops) and does obviously not admit a scalar quantum walk.
Consequently, it may be the case that, 
in addition to the special case $J(4,2)$,
most Johnson graphs with large $n$ and $k$ admit a scalar quantum walk.
\end{itemize}

\section{Conclusion}
We have obtained a simple necessary condition for the existence of a scalar
quantum walk on Cayley graphs, as well as a general method
to construct its evolution operator (or to conclude that no such walk
exists) when the necessary condition is fulfilled. 
Even if the homogeneity of the evolution
operator is required, scalar quantum walks often exist on Cayley graphs, 
and we have presented a series of examples for finitely presented groups (see Table~\ref{table}). 


\begin{table}[htb]
\begin{center}
\begin{tabular}{|c|c|c|c|}
\hline
Generators & Relations & Existence of & Groups or graphs \\
$\Delta$ & $R$ & homogeneous SQW & \\
\hline
$x$ & $\emptyset$ & yes & Cycle graphs \\ 
$x,y$ & $xy^{-1}=yx^{-1}$ & yes & Dihedral groups $D_n$, \\
 & & & Hypercube in dim. $2$ \\
$x,y,z$ & $xy^{-1}=yz^{-1}=zx^{-1}$ & yes & Symmetric group $S_3$\\
$x,y,z$ & $xy^{-1}=zx^{-1},yz^{-1}=zy^{-1}$ & yes &  Complete graph over $4$ vertices\\ 
$x,y,z$ & $xy^{-1}=yx^{-1},xz^{-1}=zx^{-1}, yz^{-1}=zy^{-1}$ & no &
 Hypercube in dim. $3$\\
\hline
$\ a^{j \frac{n}{d}+k}\ (j=0,\dots,d-1)\ $ & $a^n=e$ & yes & Complete graph with self-loops, \\
 & & & Cycle graphs \\
$a,a^{-1},a^2,a^{-2}$ & $a^6=e$ & yes & Johnson graph $J(4,2)$ \\
\hline
\end{tabular}
\end{center}
\caption{Summary of our results.
The first two columns define the groups by giving their free
group representation $\Gamma=\langle \Delta | R \rangle$,
where $\Delta$ is the set of generators and $R$ is the set of relations between these
generators. The third column specifies whether there exists a homogeneous scalar quantum walk (SQW) on the associated Cayley graph. The last column lists groups or graphs also concerned by this result.\label{table}}
\end{table} 

We have unfortunately not been able to answer completely
the open question of deriving a general necessary and sufficient
condition for the existence of a quantum operator on a graph. 
We have shown, however, that assuming homogeneity significantly 
simplifies the problem and allows to solve it explicitly 
for a large class of Cayley graphs. For non Cayley graphs, 
there is no general procedure to assign equal
coefficients to different edges in a non-arbitrary way. However,
non-homogeneous quantum walks may also be used to construct quantum
algorithms, so that exploring
which general graphs admit scalar quantum walks
certainly deserves further investigation.

\begin{acknowledgments}
O.~L.~A. thanks D. Leemans for helpful discussions 
about the presentation of groups. O.~L.~A. and J.~R. acknowledge financial
support from the Belgian National Foundation for Scientific Research (FNRS).
This work has also been supported
by the IUAP programme of the Belgian government under grant V-18
and by the European Commission under the Integrated Project Qubit 
Applications (QAP) funded by the IST directorate as Contract Number 015848.
\end{acknowledgments}

\end{document}